# $^{18}$F-PSMA-1007 salivary gland dosimetry: Comparison between different methods for dose calculation and assessment of inter- and intra-patient variability

Daniele Pistone[1,2] [§], Silvano Gnesin[3] [§], Lucrezia Auditore[1,2], Antonio Italiano[2,4], Giuseppe Lucio Cascini[5,6,7], Ernesto Amato[1,2] [*], Francesco Cicone[5,6,7]

[1]Department of Biomedical and Dental Sciences and of Morphologic and Functional Imaging (BIOMORF), University of Messina, Messina, Italy
[2]National Institute for Nuclear Physics (INFN), Section of Catania, Catania, Italy
[3]Institute of Radiation Physics, Lausanne University Hospital and University of Lausanne, Lausanne, Switzerland
[4]MIFT Department, University of Messina, Messina, Italy
[5]Department of Experimental and Clinical Medicine, "Magna Graecia" University of Catanzaro, Catanzaro, Italy
[6]Neuroscience Research Centre, PET/MR Unit, "Magna Graecia" University of Catanzaro, Catanzaro, Italy
[7]Nuclear Medicine Unit, University Hospital "Mater Domini", Catanzaro, Italy

§ Equal contributors

*corresponding author: ernesto.amato@unime.it

**Abstract**

**Objective:** Dosimetry of salivary glands (SGs) is usually implemented adopting simplified calculation approaches and approximated geometries. Our aims were i) to compare different dosimetry methods to calculate SGs absorbed doses (ADs) following $^{18}$F-PSMA-1007 injection, and ii) to assess the AD variation across patients and single SG components.

**Approach:** Five patients with prostate cancer recurrence underwent sequential PET/CT acquisitions of the head and neck, 0.5, 2 and 4 hours after $^{18}$F-PSMA-1007 injection. Parotid and submandibular glands were segmented on low-dose CT to derive SGs volumes and masses, while PET images were used to derive Time-Integrated Activity Coefficients. Average ADs to single SG components or total SG (tSG) were calculated with the following methods: i) direct Monte Carlo (MC) simulation with GATE/GEANT4; ii) spherical model (SM) of OLINDA/EXM 2.1, adopting either patient-specific or standard ICRP89 organ masses (SMstd); iii) ellipsoidal model (EM); iv) MIRD approach with organ S-factors from OLINDA/EXM 2.1 and OpenDose collaboration, with or without contribution from cross irradiation originating outside the SGs. The maximum percent AD difference across SG components ($\delta_{max}$) and across patients ($\Delta_{max}$) were calculated.

**Main results:** Compared to MC, ADs to single SG components were significantly underestimated by all methods (average relative differences ranging between -14.5% and -30.4%). Using MC, SM and EM, $\delta_{max}$ were never below 25% (up to 113%). $\delta_{max}$ values up to 702% were obtained with SMstd. Concerning tSG, results within 10% of the MC were obtained only if cross irradiation from the remainder of the body or from the remainder of the head was accounted for. The $\Delta_{max}$ ranged between 58% and 78% across patients.

**Significance:** Simple geometrical models for SG dosimetry considerably underestimated ADs compared to MC, particularly if neglecting cross-irradiation from neighboring regions. Specific masses of single SG components should always be considered given their large intra- and inter-patient variability.

# 1. Introduction

Prostate-specific membrane antigen (PSMA) is a type II transmembrane glycoprotein having neuropeptidase and folate hydrolase activity [Carter *et al* 1996, Pinto *et al* 1996], which is expressed by prostate epithelial cell membrane as well as by other normal tissues, such as salivary glands (SGs), proximal renal tubules, brain and intestine [Liu *et al* 1997]. Radiolabelled small molecule ligands of PSMA, such as PSMA-11, PSMA-617, PSMA-1007, PSMA imaging and therapy (I&T), and others, are currently being used for theragnostic of metastatic, castration-resistant prostate cancer (PC) [Eder *et al* 2012, Afshar-Oromieh *et al* 2015, Afshar-Oromieh *et al* 2016, Weineisen *et al* 2015, Giesel *et al* 2017, Sartor *et al* 2021]. The finding of a significant endothelial expression of PSMA by tumor neovasculature has raised the interest on the use of PSMA-targeting radiolabeled probes for other malignancies in addition to PC [Fragomeni *et al* 2018, Pozzessere *et al* 2019, Matsuda *et al* 2018, Tanjore Ramanathan *et al* 2020]. With the increasing number of PSMA-targeted radionuclide therapies performed worldwide, there is a growing interest on possible dose-limiting radiation-induced toxicities to salivary and lacrimal glands [Heynickx *et al* 2021, Sjögreen Gleisner *et al* 2022]. Xerostomia is a well-documented side effect in patients receiving PSMA-targeted therapies [Kratochwil *et al* 2016, Kratochwil *et al* 2017, Taïeb *et al* 2018, Heynickx *et al* 2021], and some methods for the protection of salivary glands are being evaluated clinically [Paganelli *et al* 2020, Belli *et al* 2020]. The most popular therapeutic PSMA ligand is the PSMA-617 [Sartor *et al* 2021]. The theragnostic concept implies that the biodistribution of a therapeutic radiopharmaceutical is reliably predicted by the preliminary use of a diagnostic companion. Within a few hours after administration, this paradigm is valid for the fluorinated compound $^{18}$F-PSMA-1007, owing to its structural similarity to PSMA-617 [Giesel *et al* 2016]. Therefore, in addition to its use in staging and detection of PC recurrence, $^{18}$F-PSMA-1007 can be considered a well-suited diagnostic counterpart of PSMA-617, which may help inform the selection of patients referred for PSMA-617 therapy.

The SGs include three pairs of glands, the major being the parotids and the submandibular glands. However, standard dosimetry software only recently included this tissue as a source/target region with a realistic geometry that includes separate bilateral parotids and submandibular components, such as available in the ICRP-110 phantom [ICRP 2009] with ICRP-89 organ masses [ICRP 2002]. Furthermore, in most previously published dosimetry studies, absorbed dose (AD) estimates for SGs were based on quantitative imaging employing S-values that consider only the AD to single or multiple spherical structures without considering patient-specific gland composition, geometry and mass [Afshar-Oromieh *et al* 2016, Kratochwil *et al* 2016, Kratochwil *et al* 2017, Giesel *et al* 2017, Kratochwil *et al* 2018, Rosar *et al* 2022].

The aim of the present work was twofold. First, to compare different methods for patient-specific dosimetry of the SGs in patients undergoing $^{18}$F-PSMA-1007 positron-emission tomography/computed tomography (PET/CT). Second, to assess the AD variation across different patients and single SG components. The methods adopted for AD calculations were: i) the direct Monte Carlo (MC) simulation with GATE/GEANT4 [Jan *et al* 2004, Sarrut *et al* 2014]; ii) the spherical model of OLINDA/EXM 2.1 [Stabin and Siegel 2018] using either ICRP-89 standard masses [ICRP 2002] or patient-specific gland masses; iii) the ellipsoidal model developed by Amato et al. [Amato *et al* 2014]; iv) the organ-level MIRD formalism with OLINDA S-factors and OpenDose S-factors [Chauvin *et al* 2020].

## 2. Materials and Methods

### 2.1. Patient enrollment, PET/CT calibration, acquisition and reconstruction

The study included five consecutive male patients (median age: 69 y, range: 58-85 y) with biochemical PC recurrence referred for $^{18}$F-PSMA-1007 PET/CT at the "Mater Domini" University Hospital of Catanzaro (Italy). None of the patients had previously received radiation therapy to the head and neck region, which may have altered the function and the anatomy of the SGs. All patients underwent three quantitative PET/CT segmental acquisitions of the head and neck (2 bed positions, 5 minutes/bed), 0.5, 2 and 4 hours following the injection of 300 MBq (range: 264-329) $^{18}$F-PSMA-1007 (A.C.O.M. – Advanced Center Oncology Macerata – S.R.L. Macerata, Italy), respectively. A standard diagnostic whole-body PET/CT scan was acquired 90 minutes after radiopharmaceutical administration. All the acquisitions were performed on a GE-Healthcare Discovery ST 8 slice camera, operating in 2D mode. The PET/CT device was cross-calibrated for $^{18}$F with the local activimeter (Comecer VDC-603, Comecer S.p.A, Castel Bolognese, Italy) using a cylindrical phantom (20 cm of diameter and 18 cm of internal length, total volume 5680 mL) filled with 93 MBq of $^{18}$F, obtaining a homogeneous activity concentration of 16.4 kBq/mL. The phantom acquisition (5 min/single-bed position) and reconstruction setup were the same used for patients. The reconstruction was performed with the vendor ordered subset expectation maximization (OSEM) algorithm with 2 iterations and 30 subsets, post reconstruction Gaussian smoothing of 5 mm, image matrix $128 \times 128$, pixel spacing: $4.7 \times 4.7$ mm$^2$, slice thickness: 3.3 mm. All other pertinent corrections (normalization, dead time, activity decay, random coincidence, attenuation, and scatter corrections) were applied. The study was conducted in accordance with the ethical standards of the 1964 Declaration of Helsinki and later amendments. Written informed consent was obtained from all patients.

**2.2. Salivary gland segmentation and time-activity analysis**

For each patient, a morphologic segmentation of the SGs was performed on the first time-point low-dose CT (70 mA, 120 kV, image matrix 512×512, voxel size = 0.98 × 0.98 × 3.27 mm$^3$) by three experienced operators in consensus, including two nuclear medicine specialists (FC and GLC) and one medical physicist (SG). Volumes of interests (VOIs) for the SG components (right and left parotids, right and left submandibular) were delineated by manual segmentation using the polygonal segmentation tool of PMOD v. 3.9 (PMOD technologies, Zurich, Switzerland). A total SG VOI (tSG) was obtained by the union of the four VOI subparts. The mass of each VOI was obtained by multiplying the VOI volume by the specific density of 1.045 g·cm$^{-3}$ according to ICRP Publication 89 [ICRP 2002].

For each patient and for each acquired PET, a functional segmentation was applied to define the total activity included within the SG VOIs. Using the segment editor available in the 3D Slicer software [Kikinis et al 2014], functional SG VOIs were defined by applying a threshold level of 25% of the maximum activity concentration present in the SGs, as previously considered by [Hobbs et al 2013]. This strategy took into account the activity spill-out not included using the morphological segmentation and the possible non-optimal spatial matching of SGs' activity distribution with morphologic SGs. For each patient, normalized activities within SG VOIs for each PET acquisition time ($A(t)/A_{admin}$) were computed by dividing the total activity obtained from the functional segmentation at the considered time point by the patient administered activity, therefore generating normalized time activity curves (nTACs) for each segmented SG component and for the tSG. Time integrated activity coefficients (TIACs) were obtained applying trapezoidal time integration to nTACs between t = 0 and t = 4 h, followed by analytical mono-exponential integration considering the $^{18}$F physical decay to infinite, using MATLAB v.2019b. The $A(t)/A_{admin}$ values, together with the TIACs and with the single gland masses, are reported in Supplementary Table I. Additionally, we assessed the biological behavior in SGs by computing the percentage of the injected activity normalized by the VOI mass (%IA/g). In order to remove the physical decay component, %IA/g values were decay-corrected to the administration time.

The time-integration of activities, as well as of dose-rates obtained using MC (see following section 2.3.1), was performed at the organ/suborgan level, using average values within VOIs obtained for each time-point. Hence, coregistration of the PET scans acquired at different time-points was not needed.

## 2.3. Dosimetric methods

ADs to every SG components and to tSG were calculated by several dosimetry methods. Some of those methods provide AD estimates for each SG component (see the following 2.3.1; 2.3.2; 2.3.3), while others enable the calculation of ADs to a single tSG only, implemented in the corresponding human phantom model (see the following 2.3.4; 2.3.5).

### 2.3.1. Direct Monte Carlo simulation (MC)

We performed voxel-level, patient-specific MC simulations using GATE (Geant4 Application for Emission Tomography) version 9.1, a simulation software for medical physics applications including internal dosimetry [Jan *et al* 2004, Sarrut *et al* 2014], relying on GEANT4 version 4.10.07.p02 [Agostinelli *et al* 2003, Allison *et al* 2006, Allison *et al* 2016]. For each patient, AD-rate maps were generated for each of the three consecutive PET/CT scans. For this purpose, the CTs were imported into GATE to build voxelized computational phantoms representing the patients' bodies, adopting the *Automated Hounsfield Unit stoichiometric calibration* method to define voxels' material and density. Details on this procedure can be found in [OpenGate Collaboration 2022, Ligonnet *et al* 2021, Pistone *et al* 2022]. The HU-density calibration relation by [Schneider *et al* 2000] was employed for the density interpolation, setting a *density tolerance* of 0.01 g/cm$^3$, while the materials assigned to voxels on the basis of HU intervals are reported in Table 1.

The $^{18}$F-PSMA PETs were imported to define voxelized radioactivity sources using the *linear translator* option of GATE's *imageReader* method. This assigns a decay probability corresponding to each PET voxel through a normalized linear conversion of their values, thus producing a 3D probability density map with a total sum of 1. The $^{18}$F decays (ion type primary particles) were generated using the *G4RadioactiveDecay* module and selecting the *G4EmStandardPhysics_option4* Physics List to simulate the interactions of their emitted radiation with matter. For all the simulated particles and processes, range cuts of 0.1 mm were set for the production of secondaries within the voxelized volumes. This value was significantly smaller than the CT voxel dimensions, guaranteeing an accurate spatial sampling of the energy distributions.

For each simulation, the AD at voxel level (voxels are identified by orthogonal coordinates i, j, k), $D^{ijk}_{MC}$ (Gy), was scored on the spatial grid of the corresponding CT scan, exploiting GATE's *DoseActor* with its *MassWeighting* algorithm. A total of $2 \cdot 10^8$ primary events were used in each simulation to ensure an average standard deviation of the estimated AD at voxel level below 4% within all the SGs VOIs. The simulations exploited the CPUs of the Marconi-100 cluster of the Consorzio Interuniversitario del Nord est Italiano per il Calcolo Automatico (CINECA). It enabled the parallelization of each simulation process into 25 sub-runs of $8 \cdot 10^6$ events, resulting in an average simulation time of 6 hours (the total actual CPU time for each simulation was ~150 hours).

To obtain the dose rates at voxel level for each time point, $\dot{D}^{ijk}$ (Gy/s), the $D^{ijk}_{MC}$'s were divided by the number of simulated events, $N_e = 2 \cdot 10^8$, and multiplied by the total Activity $A_{tot}$ (Bq) in the FOV of the pertinent PET acquisition:

$$\dot{D}^{ijk} = \dot{D}_{MC}^{ijk} \frac{A_{tot}}{N_e} \qquad (1)$$

For each dose rate map, the average dose rate within the tSG and SG subparts was computed with 3D Slicer using the above described functional VOIs with the additional care of excluding voxels corresponding to air or bone on CT. This was obtained by restricting the segmentations in correspondence to HU values between -900 and 290 (see Table 1). In order to take into account the differences in size between the morphological SG VOIs and the functional ones, the average dose rates were corrected by a factor given by the ratio between the volume of the functional VOI corrected by excluding air and bone ($V_{fc}$) and the volume of the morphological VOI ($V_m$). Overall:

$$\langle \dot{D} \rangle = \langle \dot{D}^{ijk} \rangle \frac{V_{fc}}{V_m} \qquad (2)$$

The average AD in SGs for each patient, $\langle D \rangle$, was calculated integrating the average dose rates $\langle \dot{D} \rangle$ as a function of time in the same way adopted for TIACs, i.e. trapezoidal integration between t = 0 and the last acquired time point, and analytical mono-exponential integration of the tail up to infinity, considering the physical decay of $^{18}$F.

*2.3.2. Spherical models (SMstd and SM)*

The average ADs to SG subparts and to tSG were computed using the spherical model available in the OLINDA/EXM 2.1 software (HERMES Medical Solution AB, Stockholm, Sweden) [Stabin and Konijnenberg 2000, Stabin and Siegel 2018], which calculates the self-AD considering homogeneous density and activity concentration in a spherical volume. Each SG subpart was treated as an individual sphere, and TIACs corresponding to each VOI were entered separately in the code. Firstly, the ICRP-89 standard masses (parotid = 25 g;

submandibular = 12.5 g) were applied (SMstd) [ICRP 2002]. In second instance, a personalized approach was adopted (SM), adjusting the sphere mass to the specific mass of each SG components derived from the CT-based segmentation. The average AD for the tSG was calculated by summing the average AD of each of the four SG subparts weighted for their respective masses.

*2.3.3. Ellipsoidal model (EM)*

The average ADs in tSG and its subparts were estimated with the analytical model developed by Amato et al. [Amato *et al* 2011, Amato *et al* 2014], which enables the calculation of absorbed fraction and self-AD in ellipsoidal homogeneous volumes of soft tissue uniformly filled with a radionuclide activity, immersed in a homogeneous medium of the same material. Each SG component was treated as an individual ellipsoid, and the AD calculation was obtained by considering the energy emission spectra of the chosen radionuclide [Stabin and da Luz 2002], setting as input the ellipsoid's axes the density (1.045 g·cm$^{-3}$ [ICRP 2002]), the TIAC, and the injected activity. To set the ellipsoids' axes, the morphologic SGs VOIs were imported into 3D Slicer and the *Labelmap Statistics* tool was employed to deduce the three so-called Oriented Bounding Box (OBB) diameters of each SG subpart (the OBB is intended as the smallest non-axis aligned bounding box that encompasses the considered segment). Consequently, the actual axes of the ellipsoids representing the glands were deduced by multiplying the OBB diameters by a factor given by the cubic root of the ratio between the OBB derived ellipsoid volume and the morphological SG subpart volume, so that the implemented ellipsoids have the same volumes (and mass) as the corresponding morphological VOI of the SG subparts.

$$Ellips.\ axis = OBB\ diameter \cdot \sqrt[3]{\frac{OBB\ ellispoid\ volume}{morph.\ SG\ volume}} \qquad (3)$$

The axes length (a,b,c, with a < b < c), elongation (E = b/c) and flatness (F = a/c) of the implemented ellipsoids were calculated (see Supplementary Table II).

As for the SM, the average AD for the tSG was calculated via weighted average of the ADs of the four SG subparts.

*2.3.4. Organ S-Factors Olinda (S-O)*

For each patient, the average AD of the tSG was calculated following the organ-level MIRD formalism implemented by OLINDA/EXM 2.1, which uses the NURBS voxel-based adult male phantom [Segars *et al* 2001] adjusted to the ICRP-89 organ masses [Stabin and Siegel 2018]. Two different AD estimations were provided:

S-O1) The first estimate considered in input to the kinetic module of OLINDA/EXM 2.1 only the tSG TIAC applied to the adult male human model, in which the tSG mass was adjusted to the actual patient tSG mass.

S-O2) In the second estimate the TIAC of the remainder of the body was also considered in input, and the total body mass of the adult male phantom was adjusted to the actual whole body mass of the patients. The TIAC of the remainder of the body was calculated by subtracting the TIAC of the tSG from the TIAC of the whole body. The TIAC of the whole body was set as in Giesel et al. [Giesel *et al* 2017] 2.64 h for $^{18}$F, in the assumption of physical decay only, neglecting biologic wash-out.

*2.3.5. Organ S-Factors OpenDose (S-OD)*

The organ-level MIRD formalism was implemented exploiting the organ S-factors provided by the OpenDose collaboration [Chauvin *et al* 2020] for the ICRP-110 adult phantoms [ICRP 2009], publicly available at https://opendose.org/svalues. The average AD to the tSG was deduced, following three estimation approaches as detailed below. In all the cases, the adult male phantom and the $^{18}$F radionuclide were selected.

S-OD1) The first estimation considered the tSG as the only source and target region. The corresponding S-factor was multiplied for the tSG TIAC and for the injected activity of the patient and the computed average AD was multiplied for the ratio between the mass of the tSG of the OpendDose adult male phantom (ICRP-110 phantom with ICRP-89 organ masses) and the actual patient tSG mass.

S-OD2) The second estimation considered the additional AD contribution from the remainder of the body (namely, the "Total body except organ contents" organ of the ICRP-110 [ICRP 2009]) set as source. In S-OD2 the corresponding S-factor to the tSG target is multiplied by the injected activity and by the TIAC in the remainder of the body (same TIAC for the remainder of the body as used in method S-O2).

S-OD3) The third estimation did not consider the contribution from the remainder of the body. In contrast, it took into account the presence of extra gland activity only in the remainder of the head, setting as source the "residual tissue, head" organ of the ICRP-110 phantom [ICRP 2009]. The TIAC in this "residual tissue, head" was defined as follows: for each PET scan, the total normalized activity was evaluated in the region given by the subtraction of the tSG functional VOI from the whole field of view of the PET scans. These total normalized activities were thus integrated through trapezoid + physical decay tail, as explained in Sec. 2.2, obtaining the TIACs. The S-OD3 approach represented the most similar scenario to the direct MC method.

**2.4. Comparisons between different methods for AD calculations**

The average AD to the tSGs and to SG subparts obtained with the different methods ($\langle D \rangle_M$) were compared in terms of relative percent differences ε (%), taking the average AD obtained with the MC method ($\langle D \rangle_{MC}$) as the reference:

$$\varepsilon(\%) = 100 \cdot \frac{\langle D \rangle_M - \langle D \rangle_{MC}}{\langle D \rangle_{MC}} \qquad (4)$$

with *M* indicating the specific dosimetric method considered. The average ε values over the examined patient population were also calculated, together with the respective standard deviations.

The Wilcoxon signed-rank test for paired samples was used to compare AD results obtained with different methods. The level of significance was set at two-tailed p < 0.05. The statistical analysis was performed using Graph Pad Prism 8.

### 2.5. Assessment of intra- and inter-patient AD variability

A maximum percent difference ($\delta_{max}$) was defined to describe the maximum average AD variation across SG components (i.e. Parotid R, Parotid L, Submandibular R, Submandibular L) for a given dosimetric method *M*:

$$\delta_{max}(\%) = 100 \cdot \frac{\max(\langle D_{Ci} \rangle_M) - \min(\langle D_{Ci} \rangle_M)}{\min(\langle D_{Ci} \rangle_M)} \qquad (5)$$

with $\langle D_{Ci} \rangle_M$ indicating the average AD in the SG component C (with $i \neq j$) for the specific dosimetric method considered.

The same metric was applied to calculate the maximum percent difference of AD to tSG across patients $\Delta_{max}$.

$$\Delta_{max}(\%) = 100 \cdot \frac{\max(\langle D_{Pk}\rangle_M) - \min(\langle D_{Pk}\rangle_M)}{\min(\langle D_{Pk}\rangle_M)} \qquad (6)$$

with $\langle D_{Pk}\rangle_M$ indicating the average AD to the tSG of the patient *k* for a given dosimetric method *M*.

## 3. Results

The nTACs of single SG subparts are shown in Figure 1 for each patient, including a box-plot representation of the biokinetics for the total SG in terms of %IA/g (raw data are provided in Supplementary Table III). This latter plot displays the biological uptake of the tSGs, still exhibiting a monotonous increment up to the last acquired time point.

### 3.1. AD to single SG components

The average ADs (mGy/MBq) to the single SG components are shown in Table 2, along with the relative percent differences ε with respect to MC. The maximum percent intra-patient AD variation $\delta_{max}$ is also reported in Table 2. All the dosimetry methods provided significantly different results compared with MC (all $p < 0.001$, Wilcoxon signed-rank test), as well as compared to each other (all $p < 0.05$, Wilcoxon signed-rank test). Considering all SG components, average ε were: -30.5 ± 36.4% for SMstd, -14.5 ± 4.3% for SM, and -15.1 ± 4.9% for EM, respectively. ADs to single SG components showed a large intra-patient variation, which was maximal (i.e. $\delta_{max}$ up to 702%) if the SMstd method was adopted (see Table 2).

### 3.2. AD to tSG

The average ADs to the tSG of each patient, together with the corresponding ε values, are reported in Figure 2. The inter-patient maximum percent tSG AD difference $\Delta_{max}$ varied between 58% and 78% across the different dosimetry methods. Raw data are reported in Supplementary Table IV. For the methods considering only tSG self-irradiation (i.e. SMstd, SM, EM, S-O1 and S-OD1), the average ε across the patient population varied between -13.0 ± 2.8% and -20.7 ± 5.5%. The methods considering the contribution of the remainder of the body, S-O2 and S-OD2, showed average ε = -8.6 ± 5.3 % and +5.7 ± 6.7%, respectively. For method S-OD3, which takes into account the AD contribution from the remainder of the head, average ε was -7.2 ± 5.0%. Likely due to the small number of observations (i.e. = 5), the differences between the various methods and MC, in terms of AD to tSG, were not statistically significant (full data not shown). Nevertheless, a trend towards significance was shown for some comparisons, such as SM vs. MC, EM vs. MC, S-O1 vs. MC, and S-OD1 vs. MC (all *p* = 0.062, Wilcoxon signed-rank test). Figure 3 shows the activity concentration and the dose-rate maps obtained in the head and neck region of patient 1 with MC at different time points.

## 4. Discussion

The information on the AD delivered to tissues is key to establish the safety and the efficacy profiles of therapeutic radiopharmaceuticals, and to provide indications on the stochastic probability of radiation-induced diseases when using diagnostic radiopharmaceuticals. The parameters influencing the accuracy of AD calculation are the specific organ geometry and mass, the composition, the density heterogeneity and the activity distribution heterogeneity, which are all taken into account by the direct MC method [Dewaraja et al 2012, Sarrut *et al* 2014, Auditore *et al* 2019, Amato *et al* 2020]. Internal dosimetry of SGs is gaining attention as SGs are the dose-limiting organ in PSMA-targeted therapy, which is being increasingly performed worldwide [Heynickx *et al* 2021, Sjögreen Gleisner *et al* 2022]. Lacrimal glands are also of concern in PSMA therapy but, to our knowledge, xerophthalmia was observed only following the administration of PSMA ligands radiolabeled with alpha emitters [Kratochwil C *et al* 2017]. An accurate AD estimation to lacrimal glands is limited by the small size of the glands and, consequently, by the significant partial volume effect which would require specific dosimetric approaches, as developed by others [Plyku *et al* 2018]. For such reasons we did not include the lacrimal glands in the present analysis.

To our knowledge, this is the first study addressing the variations between several different methods, including MC, for the calculation of patient-specific ADs to tSGs and SG sub-components following the administration of a PSMA ligand, namely $^{18}$F-PSMA-1007. The relative differences (ε) between the MC-derived ADs to SG sub-components and the other methods implementing patient-specific organ masses were, on average, -14.6 % and -15.1% for SM and EM, respectively. In a previous work [Amato *et al* 2018], we found larger differences of about 25% between SM and MC-based $^{18}$F dosimetry for more elongated cerebral structures such as the choroid plexuses targeted by small peptidic radiopharmaceuticals [Gnesin *et al* 2017, Gnesin *et al* 2020]. Though significant, the relative difference between SM and EM in terms of AD to SG components was always below 3% (Table 2), which can be explained by the modest elongation and flatness of the ellipsoids representing the SG components (Supplementary Table II), resulting in a geometry similar to a sphere. There is one single previous work adopting a MC-based approach for SG dosimetry following the administration of the fluorinated PSMA ligand $^{18}$F-DCFPyL [Plyku *et al* 2018]. In this work, a comparison was made between MC-based dosimetry and the dosimetry obtained with a region-based approach considering the SG self-irradiation and the cross-irradiation from the brain as a target [Szabo *et al* 2015]. The authors found differences of about a factor 3 between the two methods [Plyku *et al* 2018]. However, their results are not comparable to ours, as their original region-based method is hardly reproducible, and their organ masses were not specified [Szabo *et al* 2015].

The organ mass is the parameter having the largest impact on AD calculations. To account for this, we have also calculated ADs to single SG components using the SM method applying the ICRP-89 standard gland masses (SMstd). We obtained discrepancies up to a factor 4.8 between SMstd and MC (Table 2). These findings are due to the large differences between the SG masses directly calculated on single patients and the reference values of the ICRP-89. In our study, the heterogeneity of SG masses (ranges: 16.8-38.8 g and 3-11.2 g for parotids and submandibular, respectively) reflects in a large variability of ADs between single SG components of the same patient, as well as between patients. With MC, SM and EM, maximum relative AD differences between SG components ($\delta_{max}$) were as high as 113%, and never lower than 25% (Table 2). Even higher $\delta_{max}$ ranging from 53% to 702%, were obtained with SMstd. These results suggest that the calculation of an average AD for the tSG potentially masks significant AD heterogeneities between single SG components, which may result in loss of relevant clinical information for dose-response correlations, particularly in case of therapeutic administrations of PSMA-targeted radiopharmaceuticals. The largest errors are obtained if personalized organ masses are not considered. Previously, Giesel et al. calculated the ADs to SGs in a similar-sized cohort referred for $^{18}$F-PSMA-1007, using the spherical model of OLINDA/EXM 1.1 with the ICRP-89 standard masses [Giesel *et al* 2017]. The authors reported mean AD of 0.09 mGy/MBq and 0.075 mGy/MBq for the parotids and submandibular glands, respectively. These values can be directly compared with 0.063 ± 0.013 mGy/MBq and 0.030 ± 0.015 mGy/MBq we obtained applying the SMstd. The differences between our estimates and those of Giesel et al. could arise from the different TIACs obtained for the SG subcomponents in the two cohorts. TIACs were (9.7 ± 2)E-03 h and (2.4 ± 1)E-03 h for the parotid and for the submandibular glands in this work vs. (1.4 ± 0.4)E-02 h and (6.0 ± 0.3)E-03 h in [Giesel *et al* 2017].

SM and EM, used to calculate ADs to SG components, did not consider AD deposition due to cross irradiation from neighboring structures. This explains why SM and EM systematically underestimated the ADs compared to MC. This issue was addressed by additional analyses using software including realistic SG structures like OLINDA/EXM 2.1 and OpenDose. However, all the additional software-based methods used for calculations only provide ADs to tSG. Methods S-O1 and S-OD1 included the AD contribution from the cross-irradiation between the SG components. Nevertheless, compared to SM and EM, they produced even larger underestimates of the MC-based AD to tSG (average $\varepsilon$ = -20.7%, -15.5%, -13.0%, -13.4%, for S-O1, S-OD1, SM and EM, respectively, full data in Supplementary Table IV, and Figure 2). This may be due to the loss of AD in the target (tSG) if the source volume is divided in four separate sub-volumes according to the software-implemented phantoms. This, in turn, increases the surface and the mutual distance between the radiation sources and is not compensated by the cross-irradiation effect. Methods S-O2 and S-OD2 further implemented the cross-irradiation effect from the remainder of the body. A TIAC for the remainder of the body was input in the software by assuming physical decay only (no-biological voiding). As a result, the calculated ADs to tSGs were closer to those obtained with MC, with average $\varepsilon$ = -8.6% and +5.7% for S-O2 and S-OD2, respectively. The discrepancy between these two methods can be explained by the difference of the cross-irradiation S-factors (i.e. $S_{tSG \leftarrow remainder}$) used in the two software for the adult human model, i.e. 9.13E-07 mGy.s/MBq and 1.1269E-06 mGy.s/MBq for OLINDA/EXM 2.1 and OpenDose, respectively, the last value being 23% larger than the former one. The last method, S-OD3, reproduced the most similar scenario to MC by considering a TIAC for the remainder of the head. Average $\varepsilon$ was -7.2% for S-OD3, showing that AD results within 10% of the MC can be obtained only if a distribution of the activity outside the SG is accounted for (methods S-O2, S-OD2 and S-OD3). It should be acknowledged that the accuracy of all dosimetry methods adopted in the present study, including MC, could have been further improved by considering a realistic activity distribution within the remainder of the body. This was not possible as no sequential whole-body imaging was acquired, which represented a limitation of the present analysis.

A last comment deserves to be made on the observed relevant inter-patient variability ($\Delta_{max}$) of AD to tSG. With some heterogeneity between the different calculation methods, the $\Delta_{max}$ ranged between 58% and 78%, while the tSG masses showed relative differences as high as 62%, ranging between 50.7 g and 82.2 g. This further highlights the importance of patient-specific SG dosimetry for accurate AD calculation and realistic dose-response correlations. Future developments of the personalized dosimetric approach should include dose estimations in the therapeutic setting.

## 5. Conclusion

Simple geometrical models for SG dosimetry considerably underestimated ADs compared to MC, particularly if the cross-irradiation contribution from the remainder of the body or from the remainder of the head was not considered. Specific masses of single SG components, such as parotids and submandibular glands, should always be input in the computational models, given their large intra and inter-patient variability. The substitution of specific organ masses with the ICRP-89 standard reference masses produced the largest AD underestimations. On the other hand, the computational models implementing anthropomorphic voxelized phantoms which are available in newer dosimetry software such as OLINDA/EXM 2.1 or OpenDose, do not allow for the calculation of ADs to single SG components. Given the relevance of SG dosimetry in PSMA-targeting therapies, further implementation of these software to allow personalization at the level of single SG components would be welcome.

**Acknowledgments**:

We acknowledge the CINECA award under the ISCRA initiative, for the availability of high performance computing resources and support.

**Figures**

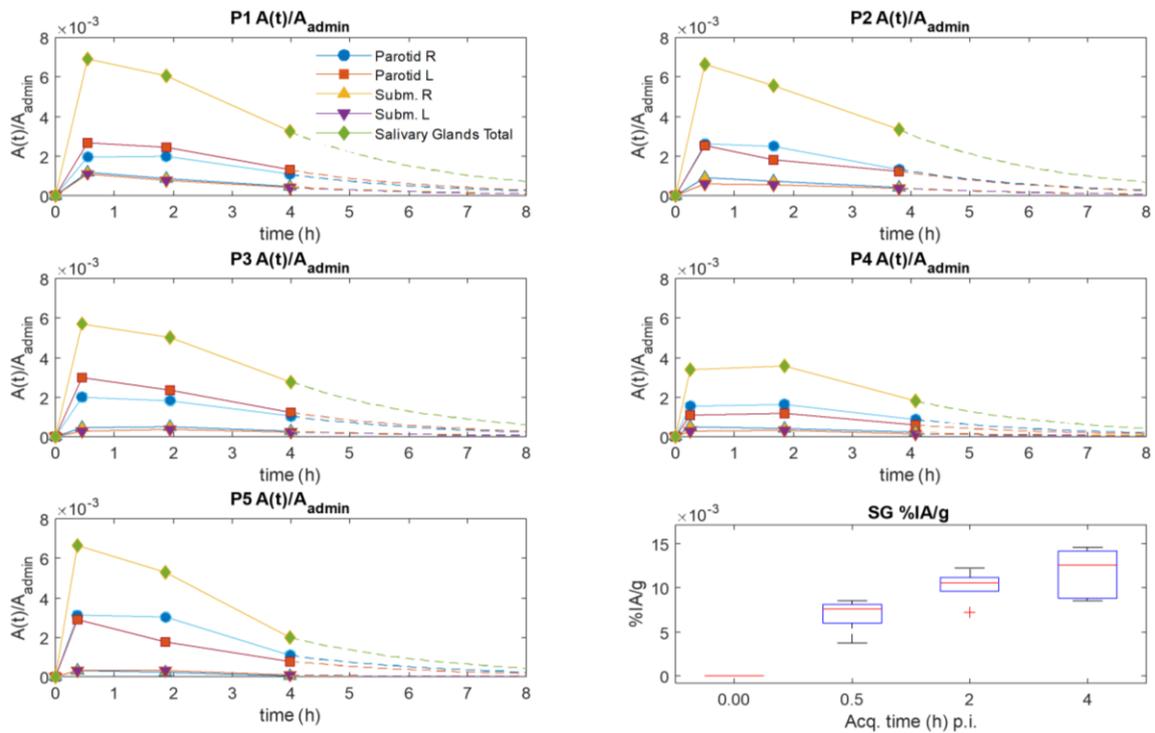

**Figure 1.** Patients' nTACs for the SG components and for tSGs. Full lines delineate TAC intervals for which trapezoidal integration was performed. Dashed lines show the extrapolated TAC considering $^{18}$F physical decay to infinity for which analytical integration was performed. The lower-right panel shows a box plot representation of the tSG biokinetic in terms of %IA/g corrected for physical decay to the administration time.

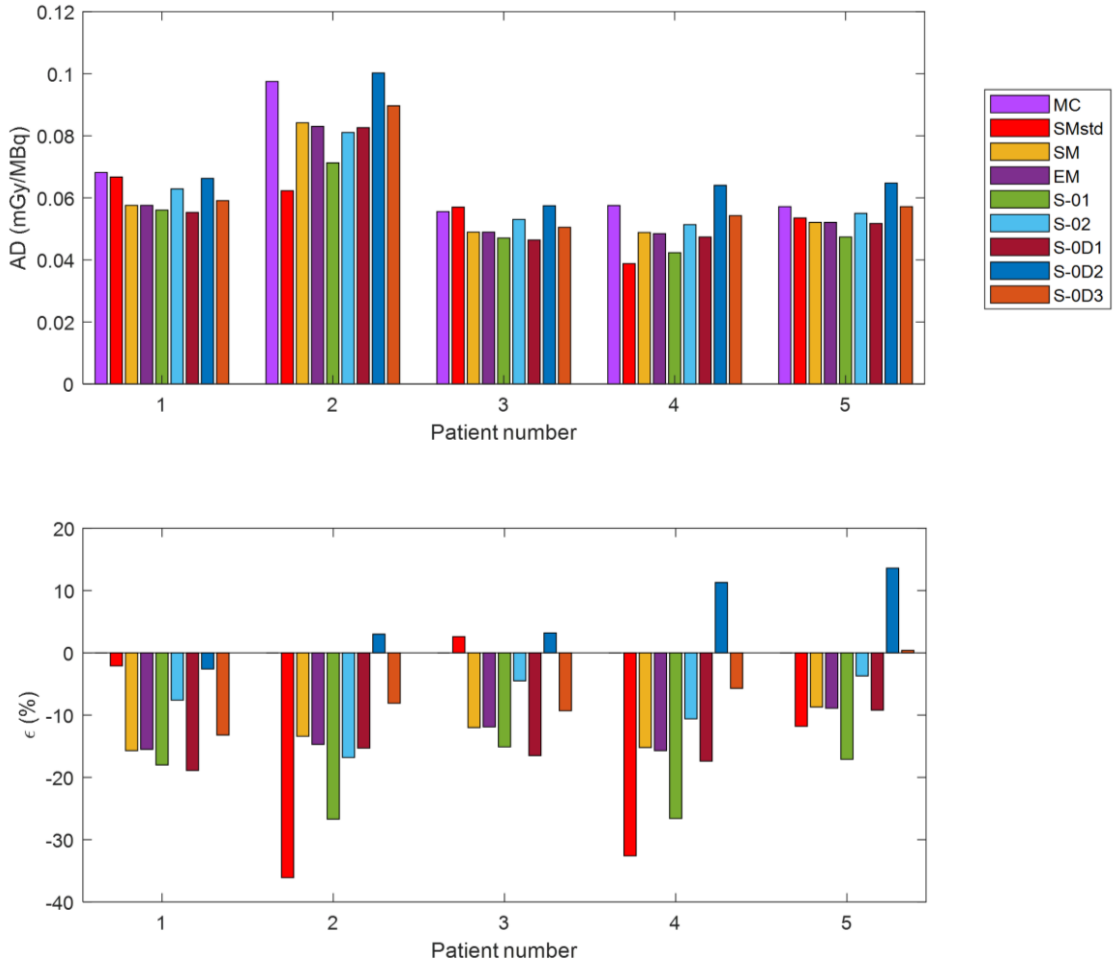

**Figure 2**. ADs to tSGs calculated with all methods described (upper panel), and corresponding relative percent difference (ε) with MC (lower panel).

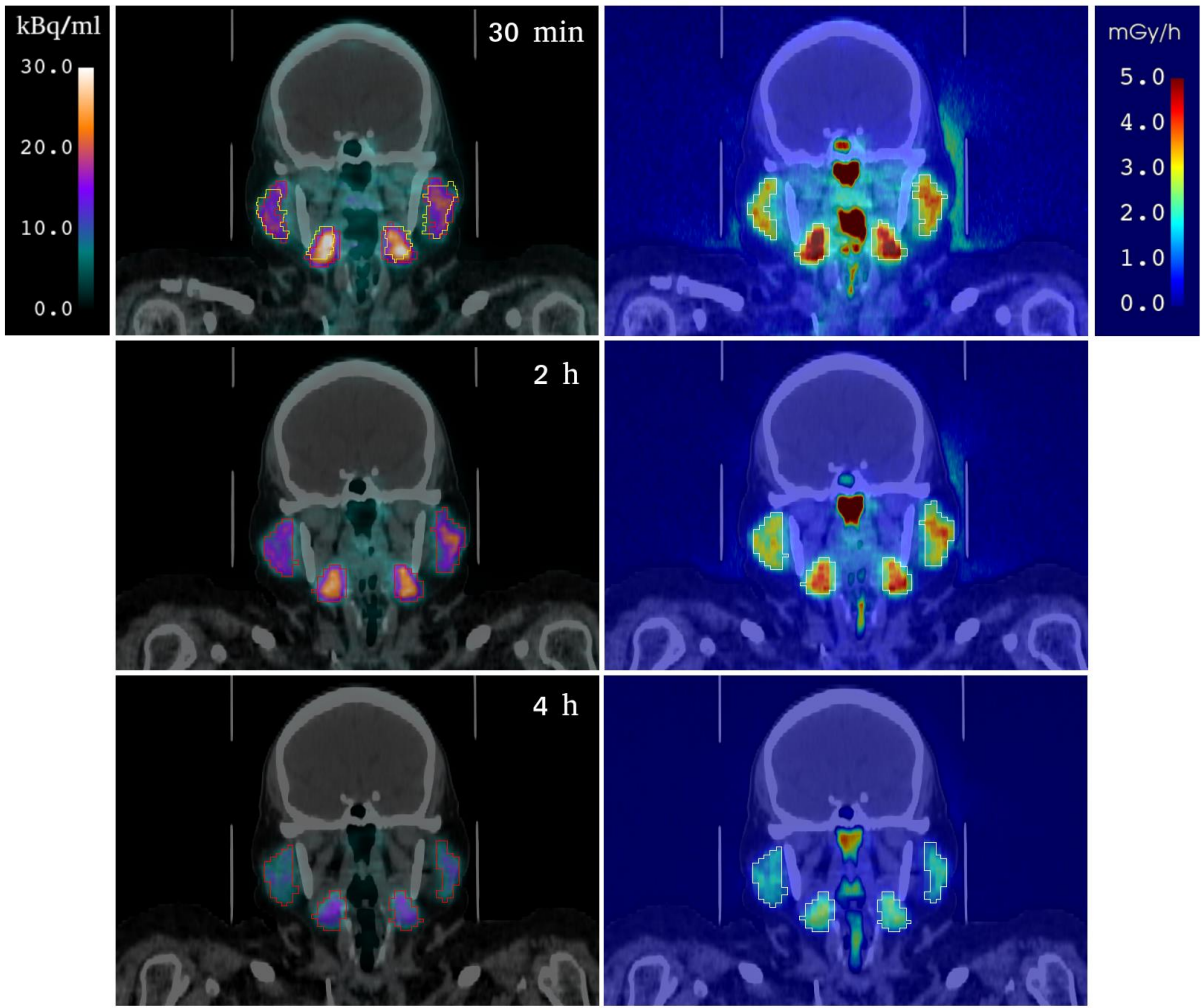

**Figure 3.** Left column: Fused PET/CT coronal slices of patient 1 showing the activity concentration in single SG components at different time points. Right column: corresponding dose rates obtained with MC simulations. The figure highlights intra and inter SGs heterogeneities of activity concentration and corresponding dose rates.

**Tables**

Table 1 HU intervals and corresponding density (ρ) intervals - through the calibration by [Schneider *et al* 2000] - set to assign materials [Geant4 Collaboration 2022] to the voxels of the patients' phantoms.

| material | HU interval | density interval ρ (g/cm³) |
|---|---|---|
| G4_AIR | HU ≤ -900 | ρ ≤ 0.10 |
| G4_LUNG_ICRP | -900 < HU ≤ -150 | 0.10 < ρ ≤ 0.85 |
| G4_ADIPOSE_TISSUE_ICRP | -150 < HU ≤ -50 | 0.85 < ρ ≤ 0.95 |
| G4_TISSUE_SOFT_ICRP | -50 < HU ≤ 290 | 0.95 < ρ ≤ 1.15 |
| G4_BONE_CORTICAL_ICRP | HU > 290 | ρ > 1.15 |

Table 2. Average ADs (mGy/MBq) obtained with the methods that allowed dose calculations for single SG subparts (MC, SMstd, SM and EM). AD relative percent difference (ε (%)) with respect to MC and maximal percent difference across SG components ($\delta_{max}$(%)) are also reported.

| | Method: | VOI Mass (g) | Dose (mGy/MBq) | | | | ε (%) | | |
|---|---|---|---|---|---|---|---|---|---|
| | | | MC | SMstd | SM | EM | SMstd | SM | EM |
| | Parotid R | 30.0 | 5.88E-02 | 5.94E-02 | 5.00E-02 | 5.01E-02 | 1.0 | -15.0 | -14.8 |
| P1 | Parotid L | 28.7 | 7.74E-02 | 7.44E-02 | 6.53E-02 | 6.56E-02 | -3.8 | -15.6 | -15.2 |
| | Subm. R | 10.9 | 7.27E-02 | 5.36E-02 | 6.11E-02 | 6.10E-02 | -26.3 | -15.9 | -16.1 |
| | Subm. L | 11.2 | 6.53E-02 | 4.85E-02 | 5.38E-02 | 5.37E-02 | -25.7 | -17.6 | -17.7 |
| | $\delta_{max}$ (%) | | 32 | 53 | 30 | 31 | | | |
| P2 | Parotid R | 16.8 | 1.23E-01 | 7.24E-02 | 1.06E-01 | 1.04E-01 | -41.1 | -13.7 | -15.0 |
| | Parotid L | 24.1 | 7.34E-02 | 6.25E-02 | 6.48E-02 | 6.39E-02 | -14.9 | -11.7 | -12.9 |
| | Subm. R | 6.6 | 9.23E-02 | 4.27E-02 | 7.87E-02 | 7.83E-02 | -53.8 | -14.8 | -15.2 |
| | Subm. L | 3.1 | 1.57E-01 | 3.42E-02 | 1.30E-01 | 1.27E-01 | -78.2 | -16.9 | -19.1 |
| | $\delta_{max}$ (%) | | 113 | 112 | 101 | 98 | | | |
| P3 | Parotid R | 32.3 | 5.11E-02 | 5.82E-02 | 4.57E-02 | 4.58E-02 | 13.9 | -10.6 | -10.4 |
| | Parotid L | 35.8 | 6.04E-02 | 7.50E-02 | 5.35E-02 | 5.35E-02 | 24.1 | -11.5 | -11.4 |
| | Subm. R | 7.6 | 5.80E-02 | 2.99E-02 | 4.79E-02 | 4.76E-02 | -48.5 | -17.5 | -18.0 |
| | Subm. L | 6.5 | 4.82E-02 | 2.20E-02 | 4.10E-02 | 4.08E-02 | -54.3 | -14.9 | -15.4 |
| | $\delta_{max}$ (%) | | 25 | 241 | 30 | 32 | | | |
| P4 | Parotid R | 24.2 | 6.04E-02 | 5.06E-02 | 5.22E-02 | 5.19E-02 | -16.2 | -13.6 | -14.1 |
| | Parotid L | 21.2 | 4.93E-02 | 3.57E-02 | 4.18E-02 | 4.17E-02 | -27.5 | -15.2 | -15.4 |
| | Subm. R | 6.2 | 6.12E-02 | 2.54E-02 | 4.97E-02 | 4.93E-02 | -58.5 | -18.9 | -19.5 |
| | Subm. L | 3.0 | 8.59E-02 | 1.79E-02 | 7.02E-02 | 6.83E-02 | -79.2 | -18.3 | -20.5 |
| | $\delta_{max}$ (%) | | 74 | 183 | 69 | 63 | | | |
| | Parotid R | 34.3 | 6.36E-02 | 8.02E-02 | 5.95E-02 | 5.94E-02 | 26.0 | -6.5 | -6.7 |
| P5 | Parotid L | 24.3 | 6.53E-02 | 5.70E-02 | 5.86E-02 | 5.87E-02 | -12.7 | -10.2 | -10.1 |
| | Subm. R | 3.6 | 4.03E-02 | 1.00E-02 | 3.29E-02 | 3.23E-02 | -75.2 | -18.3 | -19.9 |
| | Subm. L | 5.7 | 3.62E-02 | 1.46E-02 | 3.10E-02 | 3.07E-02 | -59.7 | -14.3 | -15.3 |
| | $\delta_{max}$ (%) | | 81 | 702 | 92 | 93 | | | |